\documentclass[a4paper,11pt]{article}
\usepackage{pos}

\newcommand \mres {m_{\mathrm{res}}}

\newcommand \barpsi {\langle\bar{\psi}\psi\rangle}

\title{Exploring the QCD phase diagram with three flavors of M\"{o}bius domain wall fermions}

\author*[a]{Yu Zhang}
\author[a]{Yasumichi Aoki}
\author[b,c]{Shoji Hashimoto}
\author[a]{Issaku Kanamori}
\author[b,c,d]{Takashi Kaneko}
\author[a]{Yoshifumi Nakamura}

\affiliation[a]{RIKEN Center for Computational Science,  7-1-26
	\\	Minatojima-minami-machi, Chuo-ku, Kobe, Hyogo 650-0047, Japan}

\affiliation[b]{High Energy Accelerator Research Organization (KEK), Tsukuba 305-0801, Japan}

\affiliation[c]{School of High Energy Accelerator Science, The Graduate University for Advanced Studies (Sokendai), Tsukuba 305-0801, Japan}

\affiliation[d]{Kobayashi-Maskawa Institute for the Origin of Particles and the Universe, Nagoya University, Nagoya 464–8602, Japan}

\emailAdd{yu.zhang.ey@riken.jp}

\abstract{We present an update on the study of the QCD phase transition with 3 flavors of Möbius domain wall fermions at zero baryon density. We performed simulations on lattices of size $36^3\times12\times16$ and $24^3\times12\times32$ with a variety of quark masses at a fixed lattice spacing $a=0.1361(20)$~fm, which correspond to a temperature 121(2) MeV. By analyzing 
	the chiral condensate, chiral susceptibilitities and Binder cumulant on $36^3\times12\times16$ lattices together with the result obtained from our previous study on $24^3\times12\times16$ lattices~\cite{Zhang:2022kzb}, we identified a crossover occurring at quark mass around $m_q^{\mathrm{\overline {MS}}}(2\, \mathrm{GeV}) \sim 3-4$ MeV for this temperature. 
	Besides, we show the effects of residual chiral symmetry breaking on chiral condensate and chiral susceptibilities between $L_s=16$ and 32. }

\FullConference{The 40th International Symposium on Lattice Field Theory (Lattice 2023)\\
July 31st - August 4th, 2023\\
Fermi National Accelerator Laboratory\\}

\begin{document}
{\hfill \noindent\Large KEK-CP-0396}
\maketitle

\section{Introduction}
The QCD chiral phase transition for three degenerate quark flavors,  
on the lower left corner of the Columbia plot, is yet to be determined. Based on the renormalization group flow of 3D linear sigma model, Pisarski and Wilczek predicate a first-order transition in the $N_f=3$ chiral limit~\cite{PhysRevD.29.338}. If this is the case, the first-order phase transition weakens moving away from the chiral limit and finally terminates at a critical point
of a second-order transition belonging to the 3D $Z(2)$ Ising universality class. On the contrary, a recent study of the RG flow of all couplings up to $\phi^6$ in 3D Ginzburg-Landau theory claims a possible second order phase transition in the $N_f = 3$ chiral limit ~\cite{PhysRevD.105.L071506}. Whether the first-order region exists in the $N_f=3$ light quark regime at all is still an open question. If the first-order region exists, then what is the value of the critical mass? Answering those questions requires nonperturbative lattice QCD simulations.

It is quite difficult to pin down the location of critical mass or to determine the order of phase transition in the $N_f=3$ chiral regime using lattice QCD simulations, as the computational cost increases significantly when simulating smaller quark masses. Previous studies using Wilson and staggered, or their improved fermion actions, showed that the bound of critical pion mass decreases towards finer lattices, and the critical pion mass obtained from  Wilson-type fermions differs significantly from those obtained from staggered ones~\cite{KARSCH200141,DEFORCRAND2003170, PhysRevD.54.7010, PhysRevD.91.014508,PhysRevD.96.034523,PhysRevD.95.074505}. This indicates strong cutoff and discretization scheme dependence. A recent study with unimproved staggered fermions 
suggests a second-order chiral phase transition for $N_f=3$ massless quark flavors~\cite{PhysRevD.105.034510}. Later, a study using the Highly Improved Staggered Quark action also did not find evidence of a phase transiton in pion mass range between 80 and 140 MeV~\cite{Cuteri:2021ikv}. This leaves little room for a first order region to exist, but it cannot be ruled out. The current existing results are either from staggered or Wilson type fermions which break chiral symmetry partially or entirely. Therefore, it is important to study using a chiral fermion formulation, which preserves chiral symmetry on the lattice, to further investigate this problem. We use the M\"{o}bius domain wall fermion to study the $N_f=3$ chiral region in this work. 
 
 \section{Lattice Setup}
Using the tree-level improved Symanzik gauge action and M\"{o}bius domain wall fermion action~\cite{Nakamura:2022abk}, we conducted $N_f=3$ QCD simulations using the Grid code set optimized for the Fugaku CPU A64FX~\cite{Meyer:2019gbz}. We fixed the gauge coupling at $\beta=4.0$, corresponding to a lattice spacing of $a=0.1361(20)$ fm. This value was determined from the Wilson flow $t_0$, employing the result of continuum- extrapolated $t_0$ at physical quark mass for $N_f=2+1$ QCD as  input~\cite{Borsanyi:2012zs}. In our previous study~\cite{Zhang:2022kzb}, we conducted simulations on $24^3\times 12 \times 16$ lattices with 26 bare quark masses ranging from $-$0.006 to 0.1. For this work, we performed a large volume simulation on $36^3 \times 12 \times 16$ lattices with 7 bare quark masses within the range of $[-0.005, 0.001]$, aiming to study the volume effect. Additionally, for studying the resdiual chiral symmetry breaking effect, we performed simulations on $24^3 \times 12 \times 32$ lattices with 5 bare quark masses ranging from $-$0.001 to 0.003. Each parameter set involved approximately 20,000 trajectories, and measurements were taken at every 10th trajectory.

\section{Numerical Results}

\subsection{Chiral observables}
The chiral condensate is a measure of chiral symmetry breaking in QCD. It vanishes in the chirally symmetric phase at temperature $T$ higher than the critical termperature $T_c$, retaining a nonzero value at $T< T_c$, which is the chiral symmetry broken phase. The chiral condensate is defined as a derivative of the
 partition function $Z$ with respect to the quark mass,
\begin{align}\label{eq:pbp}
	\barpsi = \frac{T}{V}\frac{\partial \ln Z}{\partial m_q} = \frac{N_f}{N_{s}^3N_{t}}\left\langle \mathrm{Tr} M^{-1}\right\rangle\,,
\end{align}
Where $M$ denotes the Dirac matrix, and $N_f$ represents the number of degenerate quark masses.
At finite quark mass, the chiral condensate diverges both additively and multiplicatively in the continuum limit. The former arises from the power-law divergence, which is proportional to
$m_q a^{-2}$. For domain wall fermions with finite $L_s$, the leading effect of 
small but finite violation of chiral symmetry 
causes an additive renormalziation to the bare input quark mass, known as the residual mass $m_{\rm{res}}$. There is an additional power divergence induced by this residual chiral symmetry breaking effect, which is proportional to $x\mres a^{-2}$ with an unknown coefficient $x$~\cite{Sharpe:2007yd}. 
Therefore, the chiral condensate for domain wall fermions is expected to behave as
\begin{align}\label{eq:pbp}
	\barpsi|_{\mathrm{DWF}} \sim C\frac{m_q + x\mres}{a^2} + \barpsi|_{\mathrm{cont}} + ...\,.
\end{align}
In the chiral limit $m_q + \mres = 0$, there still remains some UV divergent pieces, represented as $C\frac{(x-1)\mres} {a^{2}}$. One way of eliminating the additive divergence is to use the finite temperature chiral condensate $\barpsi^{T>0}$ minus the zero tempearture chiral condensate $\barpsi^{T=0}$ at the same quark mass, as this divergence term ($C\frac{m_q + x\mres}{a^2}$ ) remains temperature independent. 
To address the remaining multiplicative logarithmic divergence, we use the bare chiral condensate divided by the quark mass renormalization constant $Z_m$ in the 
$\overline{\rm{MS}}$ scheme at a scale of $\mu = 2$ GeV. $Z_m^{\overline{\rm{MS}}}(2\,\rm{GeV})$ is determined using an inter/extra-polation as a function of $\beta$ with those determined in a non-perturbative renormalization~\cite{Tomii:2016xiv}. See a description in Ref.\cite{Aoki:2021kbh} for detail.
Then the renormalized chiral condensate can be defined as
\begin{align}\label{eq:renorm_pbp}
	[\barpsi^{\mathrm{T}>0} - \barpsi^{\mathrm{T}=0}]^{\overline{\mathrm {MS}}}(2\,\mathrm{GeV}) = \frac{\barpsi^{\mathrm{T}>0} - \barpsi^{\mathrm{T}=0}}{Z_m^{\overline{\mathrm {MS}}}(2\,\mathrm{GeV})}\,.
\end{align}

For the determination of the transition point, we use the disconnected chiral susceptibility $\chi_{\rm{disc}}$, which describes the fluctuation of the chiral condensate.  For $N_f$ quark flavors, it is given by
\begin{align}\label{eq:chi_disc}
	\chi_{{\rm disc}} = \frac{N_f^2}{N_s^3 N_t}\Bigl(\left\langle ({\rm Tr} M^{-1})^2 \right\rangle - \left\langle {\rm Tr} M^{-1} \right\rangle^2 \Bigr)\,.
\end{align}
$\chi_{\rm{disc}}$ is a part of the total chiral susceptibility $\chi_{\rm{\sigma}}$, which is defined as the derivative of $\barpsi$ with respect to the quark mass. The other part is the connected chiral susceptibility,
\begin{align}\label{eq:chi_tot}
\chi_{\rm{\sigma}} = \frac{\partial \barpsi}{\partial m_q} =  	\chi_{{\rm disc}} + \chi_{\rm {con}}\,,
\end{align}
\begin{align}\label{eq:chi_con}
\chi_{{\rm con}} = -\frac{N_f}{N_s^3 N_t} \left\langle {\rm Tr} M^{-2} \right\rangle\,.
\end{align}
We will show the results of both disconnected chiral susceptibility and total chiral susceptibility with only multiplicative renormalization applied.
 Although the total chiral susceptibility suffers from an additive 
 divergence, it is constant as a function of quark mass thus has no 
 effect for the determination of the transition point.



\begin{figure}[!htp]
	\centering
	\includegraphics[width=0.45\textwidth, height=0.215\textheight]{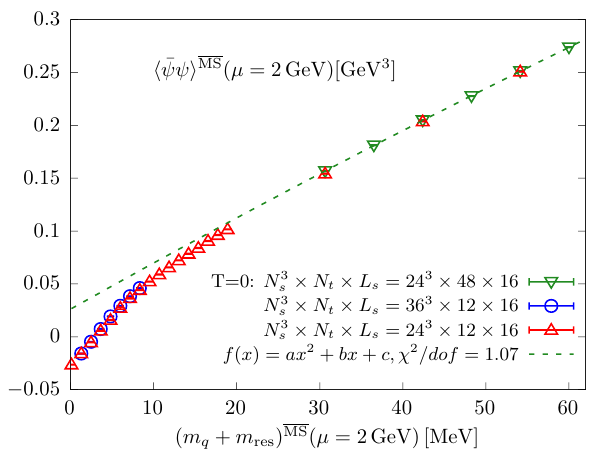}
	\includegraphics[width=0.45\textwidth, height=0.215\textheight]{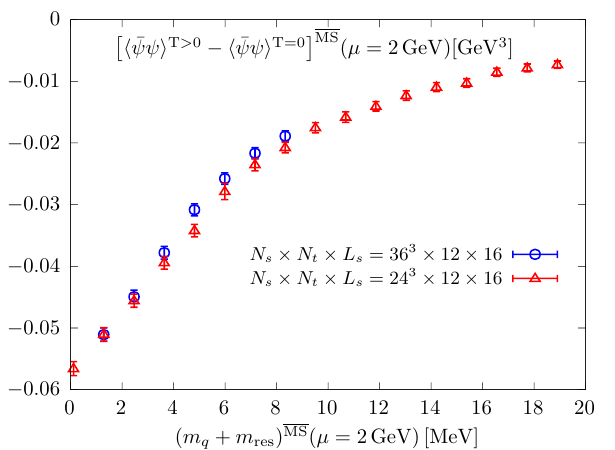}
	\caption{Left: The multiplicatively renormalized chiral condensate on finite temperature ensembles  $N_s \times12 \times 16$ with  $N_s=24, 36$ and zero temperature ensembles $24^3\times48 \times 16$ as a function of quark mass for $\beta=4.0$. The dashed line represents the quadratic extrapolation for zero temperature chiral condensate. Right:  The renormalzied subtracted chiral condensate as a function of quark mass for $N_s \times12 \times 16$ lattices with  $N_s=24$ and $36$.} 
	\label{fig:Nt12_pbp}
\end{figure}

On the left panel of~\autoref{fig:Nt12_pbp}, we show the multiplicatively renormalized chiral condensate as a function of the renormalized quark mass for finite temperature $N_t=12$ lattices alongside zero temperature lattices. The negative finite temperature result near the chiral limit arises from the remaining UV divergent piece $C\frac{(x-1)\mres} {a^{2}}$.
To eliminate the UV divergence term $C\frac{m_q + x\mres}{a^2}$, we utilize $\left \barpsi^{\mathrm{T}>0} - \barpsi^{\mathrm{T}=0} \right \vert_{m_q + \mres}$. The zero temperature results are derived from the green dashed line representing the quadratic extrapolation based on existing zero-temperature data.
Then the renormalized subtracted chiral condensate  
is depicted on the right panel of~\autoref{fig:Nt12_pbp}. We observe a small finite volume effect, a difference between $24^3$ and $36^3$ lattices, at $(m_q + \mres)^{\overline{\mathrm {MS}}}(2\,\mathrm{GeV}) \sim 4-9$ MeV and a rapid transition at $(m_q + \mres)^{\overline{\mathrm {MS}}}(2\,\mathrm{GeV}) \sim 1-5$ MeV. 
To precisely locate the inflection point, we focus on the chiral susceptibilities, which are more sensitive observables for studying the phase transition as they measure the response of the chiral condensate to a small change in the quark mass.

\begin{figure}[!htp]
	\centering
	\includegraphics[width=0.45\textwidth, height=0.215\textheight]{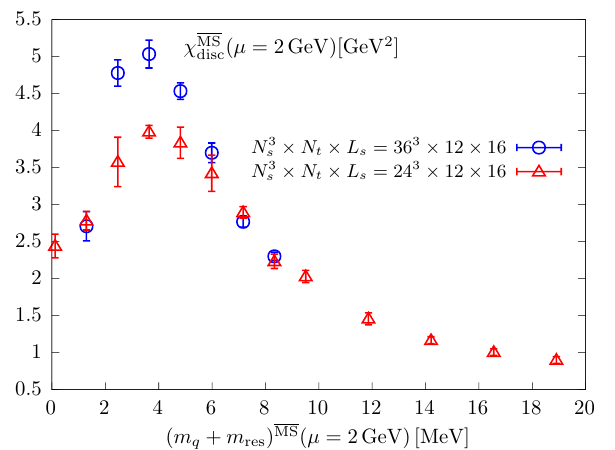}
	\includegraphics[width=0.45\textwidth, height=0.215\textheight]{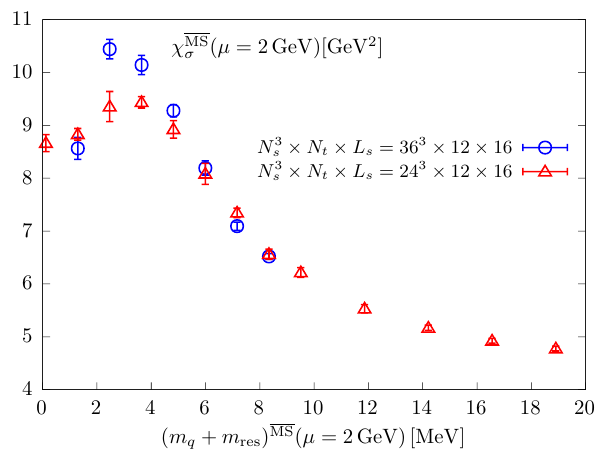}
	\caption{The renormalized disconnected chiral susceptibility (left) and the multiplicatively renormalized total chiral susceptibility (right) 
		as a function of quark mass for $N_t=12$ lattices at $\beta=4.0$.} 
	%
	\label{fig:Nt12_chi_sus}
\end{figure}

On the left panel of~\autoref{fig:Nt12_chi_sus}, we present the disconnected chiral susceptibility as a function of quark mass for $N_t=12$ lattices with two different volumes. The observed transition mass point is around 4 MeV. We notice a sizeable finite volume effect in the vicinity of the transition range, approximately $(m_q + \mres)^{\overline{\mathrm {MS}}}(2\,\mathrm{GeV}) \sim 2-5$ MeV. On the right panel of~\autoref{fig:Nt12_chi_sus}, we observe a relatively small but significant finite volume effect from the total chiral susceptibility within the same quark mass range. 

\begin{figure}[!htp]
	\centering
		\includegraphics[scale=0.7]{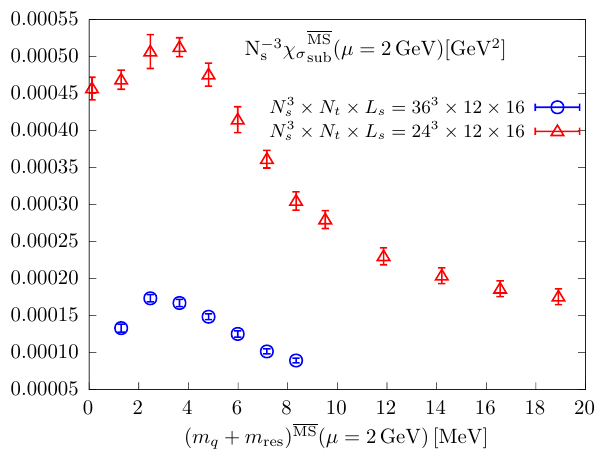}
	\includegraphics[scale=0.7]{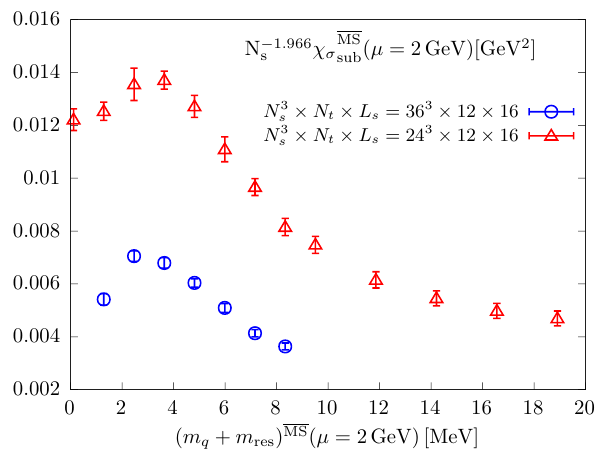}
	\caption{Rescaled subtracted total chiral susceptibility with the scaling factor corresponding to a first order phase transition (left) and  Z(2) second order phase transition (right) for $N_t=12$ lattices.} 
	\label{fig:Nt12_rescale_chi_tot}
\end{figure}

Based on finite-size scaling analysis, if it is a true phase transition, then at the critical point, the maximun of the total chiral susceptibility would be proportional to $N_s^3$ for a first-order phase transition and $N_s^{1.966}$ for a Z(2) second-order phase transition. However, the total chiral susceptibility contains a constant term due to UV power divergence, which is independent of volume and will distort the volume scaling. Therefore, we calculated the subtracted total chiral susceptibility $\chi_{\sigma_{\mathrm{sub}}}^{\mathrm{\overline{MS}}}$ with the UV divergence eliminated explicitly.
The subtracted total chiral susceptibility times inverse of the scale factor for these two possibilities is depicted in~\autoref{fig:Nt12_rescale_chi_tot}. Here, the rescaled subtracted total chiral susceptibility at the peak location significantly differs from what is expected in the case of a first-order transition or a Z(2) second order phase trasnsition. This suggests that it's most likely a crossover transition. The transition mass point estimated from the disconnected chiral susceptibility and total chiral susceptibility is around 4 MeV and 3 MeV, respectively. This discrepancy is understandable for a corrossover, where the transiton occurs over a broad range. We expect the transition mass point obtained from these two susceptibilities  will concide at the chiral phase transition termperature.

\subsection{Binder cumulant}
To investigate the nature of phase transition, we utilize the Binder cumulant of the chiral condensate $B_4(\bar\psi \psi)$, defined by~\cite{cite-key}

\begin{align}\label{eq:B4}
	B_4(\bar\psi \psi) = \frac{\left\langle (\delta \bar\psi \psi)^4\right\rangle}{\left\langle (\delta\bar\psi\psi)^2\right\rangle^2},\quad \delta\bar\psi \psi = \bar\psi \psi - \barpsi\,.
\end{align}
For the chiral critical point in $N_f=3$ QCD, $B_4(\bar\psi \psi)$ will approach a universal value of 1.604~\cite{Blote_1995} on finite volumes. However, for transitions rather than critical points, $B_4(\bar\psi \psi)$ takes a characteristic value in the infinite volume limit. Specifically, for a crossover, $B_4(\bar\psi \psi)=3$ in the infinite volume limit. For a first-order transition, the Binder cumulant at infinite volume is $B_4(\bar\psi \psi)=1$.

\begin{figure}[!htp]
	\centering
	\includegraphics[width=0.45\textwidth, height=0.215\textheight]{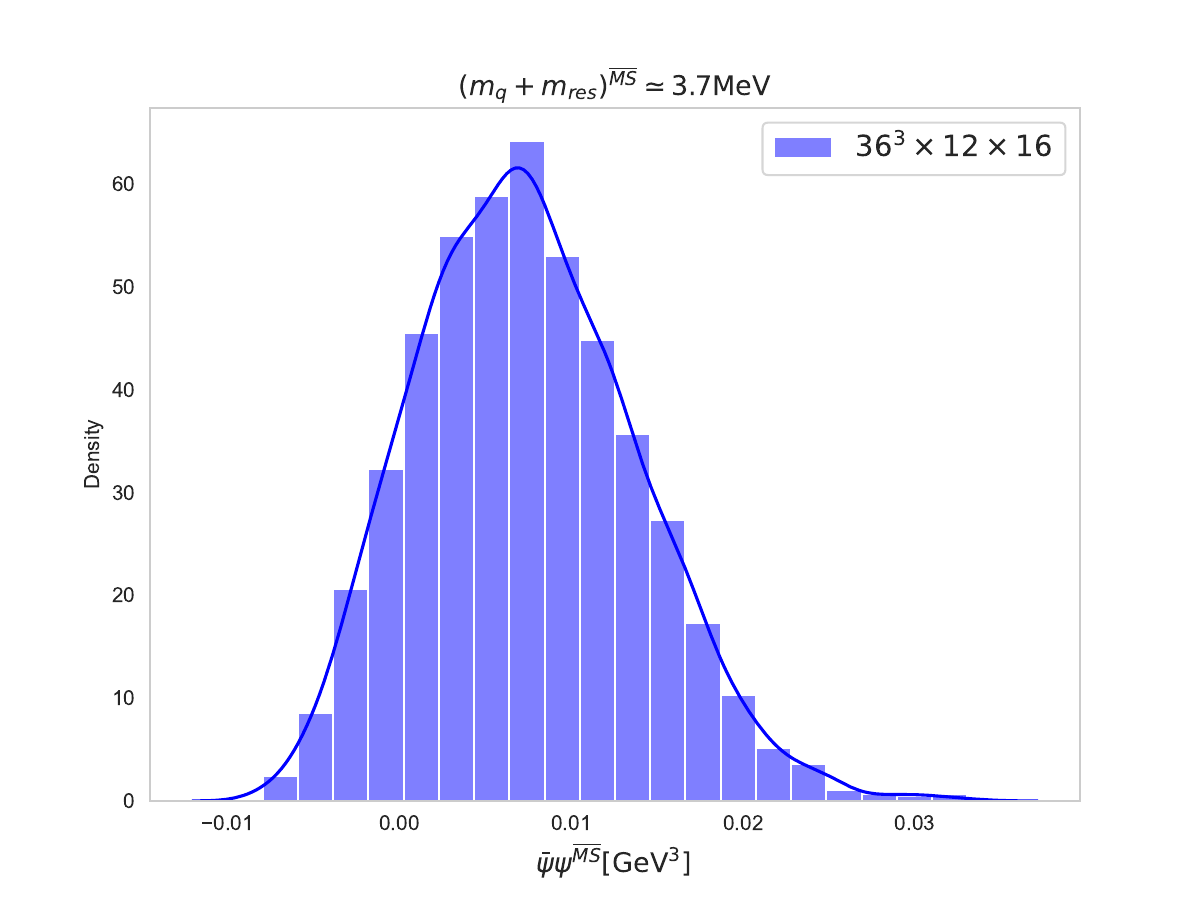}
	\includegraphics[width=0.45\textwidth, height=0.215\textheight]{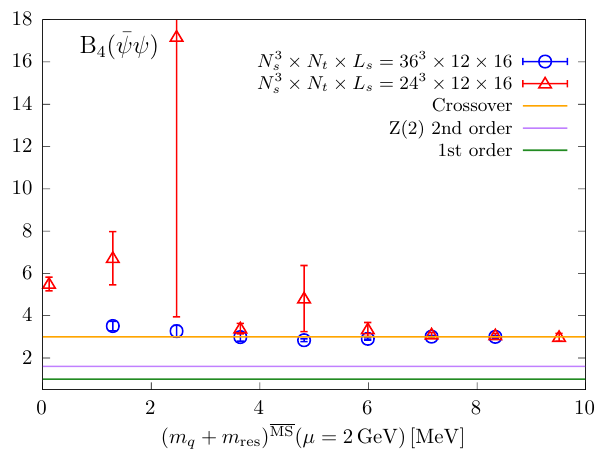}
	\caption{Left: The histogram of chiral condensate in the vincity of the transition mass point for $36^3 \times 12 \times 16$ lattice. Right: The binder cumulant as a function of quark mass for $N_t=12$ ensembles.} 
	\label{fig:Nt12_hist_B4}
\end{figure}

Before presenting the result of Binder cumulant, we display the distribution of the chiral condensate near the transiton point for $36^3 \times 12 \times 16$ lattice, in the left panel of~\autoref{fig:Nt12_hist_B4}. We observe it behaves like a Gaussian distribution rather than a double peak structure, as expected for a first order phase transition. This provides a further evidence for a crossover transition. The right panel of~\autoref{fig:Nt12_hist_B4} illustrates $B_4(\bar\psi \psi)$ as a function of quark mass for $N_t=12$ lattice ensembles. The results for $36^3 \times 12$ lattice ensembles are consistent with 3 near the transition masses of 3 and 4 MeV, as determined from the peak of total and disconnected chiral susceptibility. For $24^3 \times 12$ lattice ensembles, the result is consistent with 3 near transition mass point 4 MeV. However, it is quite large and not consistent with 3 in one standard deviation near transition point 3 MeV but it is much far from 1.604 and 1. These indicate a crossover transition.

\subsection{Residual chiral symmetry breaking effect for different $L_s$}
The M\"obius domain wall fermion with finite $L_s$ introduces a small but non-negligible amount of chiral symmetry breaking. To the leading order in an expansion in lattice spacing, this residual chiral symmetry breaking can be characterized by the residual mass $m_{\rm{res}}$, which acts as an additive shift to the bare input quark mass, resulting in the total quark mass $m = m_q + m_{\rm {res}}$. One way of simulating the small value of target quark mass is to keep the value of  $m_{\rm{res}}$ small, which can be achieved by increasing $L_s$.

\begin{figure}[!htp]
	\centering
	\includegraphics[scale=0.6]{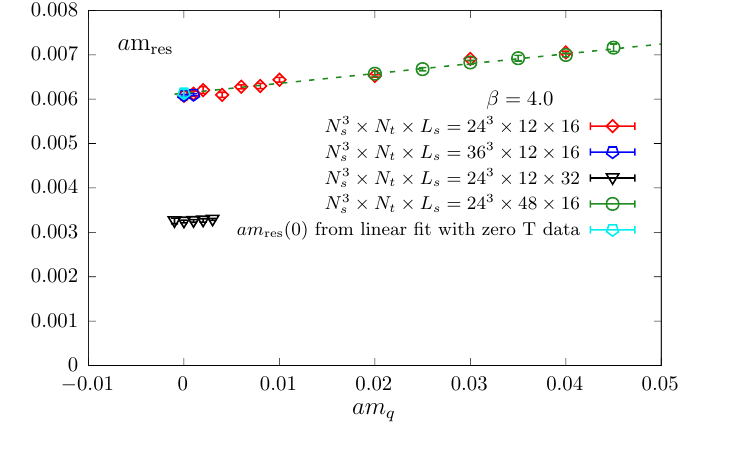}
	\includegraphics[width=0.45\textwidth, height=0.215\textheight]{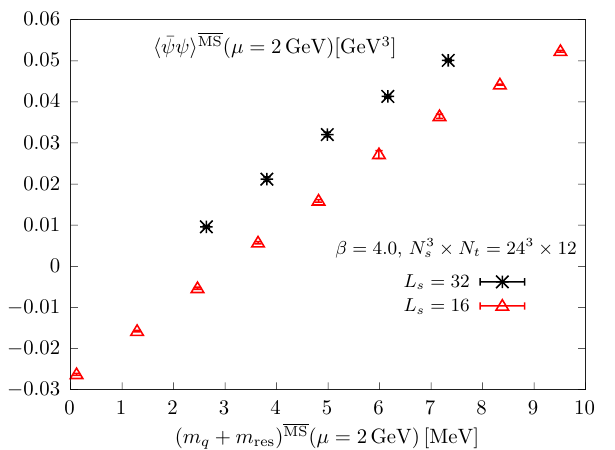}
	\caption{Left: The residual mass as a function of bare input quark mass for zero temperature and finite temperature ensembles with different $L_s$ at $\beta=4.0$. The dashed line denotes the linear fit to the zero temperature result. Right: chiral condensate as a function of quark mass for $24^3 \times 12 \times L_s$ lattices with $L_s=16$ and $32$.} 
	\label{fig:Nt12_mres_pbp_Ls32}
\end{figure}

The left panel of~\autoref{fig:Nt12_mres_pbp_Ls32} displays the residual mass as a function of bare input quark mass for zero temperature ensembles with $24^3 \times 48 \times 16$ lattice and finite temperature ensembles with $L_s=16$ and $32$ at $\beta=4.0$. The green dashed line represents the linear fit to the zero temperature data. 
The common method for defining a mass-independent $m_{\rm{res}}$ involves an extrapolation to the zero input quark mass limit, which is determined to be 0.00613(9) for $L_s=16$ and 
0.00325(3) for $L_s=32$. A $1/L_s$ dependence dominates the contribution to $m_{\rm{res}}$. We find that the $m_{\rm{res}}$ calculated from both zero temperature and finite temperature ensembles are consistent and nearly independent of the volume, see also~\cite{Jung:2000fh}.

The right panel of~\autoref{fig:Nt12_mres_pbp_Ls32} shows the chiral condensate as a function of quark mass for $24^3 \times 12 \times L_s$ lattices with $L_s=16$ and $32$. A negative result near the chiral limit is observed for $24^3 \times 12 \times 16$ lattices, attributed to the remaining UV divergence term $C\frac{(x-1)\mres} {a^{2}}$, as previously mentioned, indicating $C(x-1)<0$. An increase of the chiral condensate is observed when increasing $L_s$ from 16 to 32 while keeping the total quark mass the same. This can be easily understood from $\barpsi|_{\mathrm{DWF}} \sim C \big(\frac{(x-1 )\mres}{a^2} +  \frac{m_q +\mres}{a^2}\big) + \barpsi|_{\mathrm{cont}}+ ...\,.$, given that $m_{\rm{res}}$ for $L_s=16$ is larger than for $L_s=32$, and $C(x-1) < 0$. If we do a simple extrapolation, we will see the $\barpsi$ will have a smaller chiral symmetry breaking effect for $L_s=32$ compared to $L_s=16$ in the chiral limit, as expected.

\begin{figure}[!htp]
	\centering
	\includegraphics[width=0.45\textwidth, height=0.215\textheight]{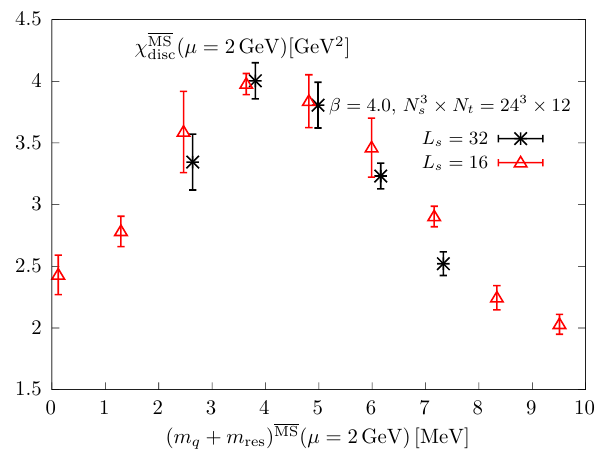}
	\includegraphics[width=0.45\textwidth, height=0.215\textheight]{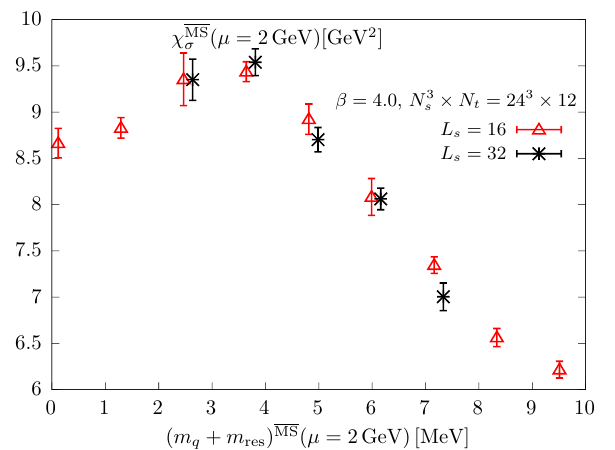}
	\caption{The disconnected chiral susceptibility (left) and total chiral susceptibility (right) as a function of quark mass for $24^3 \times 12 \times L_s$ lattices with $L_s=16$ and $32$. } 
	\label{fig:Nt12_chi_sus_Ls32}
\end{figure}

\autoref{fig:Nt12_chi_sus_Ls32} shows the result of disconnected chiral susceptibility (left) and total chiral susceptibility (right) as functions of quark mass for two different $L_s$. We expect that the $xm_{\mathrm{res}}$ effect will not show up in the susceptibility as it is only constant as we change the quark mass. In fact, with only the proper quark mass shift, the results from different $L_s$ appear consistent with each other. This is taken as evidence that we have a good control of the explicit chiral symmetry breaking already at smaller fifth size $L_s=16$, where the position of the susceptibility peak is in the negative input quark mass range.

\section{Summary and Outlook}
We carried out the study of $N_f=3$ QCD phase transition using M\"obius domain wall fermions at a fixed lattice spacing of $a=0.1361(20)$ fm on $36^3 \times 12 \times 16$ lattices, corresponding to a temperature of 121(2) MeV. We simulated various quark masses for this temperature. Through the analysis of the chiral condensate, chiral susceptibilities and Binder cumulant, comparing them with the results obtained on $24^3 \times 12 \times 16$ lattices in our previous work~\cite{Zhang:2022kzb}. We find the transition mass point is around $m_q^{\mathrm{\overline {MS}}}(2\, \mathrm{GeV}) \sim 4$ MeV and 3 MeV, as determined from the disconnected and total chiral susceptibility, respectively, at a temperature of 121(2) MeV and it is a crossover transition. Should there be a first-order region in the lower left corner of the Columbia plot, the critical quark mass would be smaller than 4 MeV. This is consistent with the findings using Wilson and staggered type fermions.

To reduce the residual chiral symmetry breaking effects, we conducted simulations on $24^3\times 12\times 32$ lattice. We found $1/L_s$ dependence dominates the contribution to the residual mass. By investigating the residual chiral symmetry breaking effect on chiral condensate and chiral susceptibilities between $L_s=16$ and 32 lattices, we found that in the chiral limit
the residual chiral symmetry breaking effect for chiral condensate  is smaller for $L_s=32$ than for 16, as expected. The chiral susceptibilities results are consistent at the approximates same total quark mass between these two different $L_s$, which indicates we have a good control of the explicit chiral symmetry breaking already at smaller $L_s=16$.

For the crossover transition, the peak height of chiral susceptibilities will remain constant for larger volumes. However, we observed some volume dependence in the chiral susceptibilities for $N_s^3 \times 12 \times 16$ lattices between $N_s=36$ and 24, though not as substantial as one would expect from a true phase transition. Presently, we are conducting simulations on larger lattice of $48^3\times 12\times 16$. Additionally, to directly probe the first order region, if it exist, we are working toward even lower temperatures through simulations on $N_t=14$ lattices, corresponding to a temperature of 104(2) MeV.

\section*{Acknowledgments}
This work used the computational resources of Supercomputer Fugaku provided by the RIKEN Center for Computational Science through HPCI project hp210032 and Usability Research ra000001 as well as Wisteria/BDEC-01 Supercomputer System at Tokyo University/JCAHPC through HPCI project hp220108 and Ito supercomputer at Kyushu University through HPCI project hp190124 and hp200050 and also the Hokusai BigWaterfall at RIKEN. This work is also supported in part by JSPS KAKENHI grant No 20H01907 and 21H01085.  We acknowledge the Grid Lattice QCD framework\footnote[1]{https://github.com/paboyle/Grid} and its extension for A64FX processors~\cite{Meyer:2019gbz} which is used to generate the QCD configurations for this study. We thank N. Meyer and T. Wettig for discussions on the use of Grid for A64FX. For the measurement, we used 
Hadrons code~\cite{antonin_portelli_2020_4293902}.


\bibliographystyle{JHEP.bst}
\bibliography{ref.bib}


\end{document}